Trapping a magnetic field of 14.8 T using stacked coated conductors of 12 mm width

Masahiro Suyama[1], Sunseng Pyon[1], Yasuhiro Iijima[2], Satoshi Awaji[3], and Tsuyoshi Tamegai[1]

[1] Department of Applied Physics, The University of Tokyo, 7-3-1 Hongo, Bunkyo-ku, Tokyo 113-8656, Japan

[2] Fujikura Ltd., 1-5-1 Kiba, Koto-ku, Tokyo 135-8512, Japan

[3] Institute for Materials Research, Tohoku University, 2-1-1 Katahira, Aoba-ku, Sendai 980-8577, Japan

**Abstract**

We fabricated a compact ($13 \times 12 \times 12.5$ mm$^3$) trapped-field magnet by stacking 200 pieces of EuBa$_2$Cu$_3$O$_7$ coated conductors with BaHfO$_3$ nanorods as artificial pinning centers. It was magnetized by field-cooling method using a 18 T superconducting magnet, and the maximum field of 14.8 T was trapped at 10 K at a field-ramp rate of 0.1 T/min. At higher temperatures, although the trapped field was decreased, we could trap the field with much faster ramp rate. In order to understand these results, we also performed calculations of the trapped field based on $J_c - H$ characteristics of the coated conductor. The calculated trapped field value was larger than the experimental value at 10 K. This discrepancy can be understood by considering the reduction of the effective size of the stacked coated conductors due to the existence of extended regions of irregular flux penetration called "flux jets".

# 1. Introduction

Cuprate superconductors (RE)Ba$_2$Cu$_3$O$_7$ (RE: Rare Earth), or REBCO, have higher critical current density than other superconductors at high temperatures and high fields. They are expected to be used for superconducting cables for power transmission and magnets that can generate high magnetic fields. Extensive developments of REBCO tapes, called coated conductors (CCs), have been conducted world-wide, and they are now commercially available [1]. Meanwhile, REBCO bulk materials have been developed, which can be used as a magnet by trapping an externally applied magnetic field [2-6]. It has been reported that a field of 17.6 T was trapped in a bulk GdBCO reinforced by a stainless steel ring [2]. Although bulk superconductors can trap a high field, they have issues related to their mechanical strengths and thermomagnetic instabilities [3,7]. In order to prevent bulk superconductors being destroyed in high magnetic fields, mechanical reinforcements such as using a stainless steel ring [2], resin impregnation, and carbon fiber wrapping [3] are necessary. As alternative candidate materials for magnets using bulk superconductors, MgB$_2$ [8,9] and iron-based superconductors (IBSs) [10] have been also studied. Unlike REBCO, since polycrystalline materials of MgB$_2$ and IBSs have reasonably high $J_c$, the bulks of these materials can be fabricated by hot pressing of their powders, while REBCO bulk is fabricated by top seeded melt growth. Hence, although the trapped field of these materials are lower than that of REBCO bulks, they have advantages in the mechanical strength and the simplicity of fabrication process. Utilizing unique electormagnetic properties of bulk superconductors, magnet lenses, which can generate higher magnetic fields than the applied field, have also been developed [11,12]. Furthermore, hybrid trapped field magnet lenses consisting of REBCO lenses and high temperature superconductor (HTS) bulk cylinders have been reported, and they can generate higher magnetic fields than the applied field without background fields [13,14]. As an alternative method to overcome the issues of REBCO bulk materials, there have been many studies on the fabrication of trapped-field magnets by stacking CCs [15-21], and a field of 17.7 T in a stack of two kinds of CCs with a total volume of 3.5×10$^4$ mm$^3$ has been realized, which is the highest among all trapped-field magnets ever reported. Among magnets fabricated from CCs with a standard width of 12 mm, 13.4 T has been the highest field ever reported [17]. Stacked magnets, which consist of stacked CCs, have great mechanical strength because a large part of CC is Hastelloy substrate layer, and also have improved thermomagnetic stability because Ag or Cu layers are used for protection and thermal stability. Therefore, stacked magnets have great potential in real applications for trapped-field magnets.

In this paper, we fabricated a stacked magnet from EuBCO CCs with embedded BaHfO$_3$ nanorods with excellent $J_c$ characteristics at low temperatures and high fields. We succeeded in trapping 14.8 T at 10 K after field cooling in 18 T magnetic field. First, we show the process of fabrication of the stacked magnet in Sec. 2. It is followed by the assessment of cutting methods of CCs in Sec. 3.1. Then, the result of magnetization and relaxation of the magnet are described in Sec. 3.2.

After that, the comparison between the experimental and calculated trapped fields and our understanding of the discrepancy between these values are discussed in Sec. 3.3. Finally, we summarize this paper in Sec.4.

## 2. Experimental Methods

In the present study, we used EuBCO CCs with $BaHfO_3$ nanorods [22,23] with 12 mm width fabricated by Fujikura Ltd., which has transport critical current 685 A at 77 K under self-field. The thickness of the CCs is about 60 μm, including 50 μm Hastelloy substrate, 2.5 μm EuBCO layer, and 2 μm Ag layer for protection and thermal stability. The CCs were cut into 13 mm length due to the reason described below, and 200 pieces of CCs were prepared. These CCs were stacked by sandwitching 0.5 mm U-shaped Cu plate between two 100 CCs stacks as shown in Fig. 1(a) to place Hall probes (HG-0711, Asahi Kasei Microdevices) for the measurements of trapped magnetic field. Therefore, the total dimensions of the stacked magnet is 13×12×12.5 $mm^3$ including the Cu plate in the middle. In the Cu plate, two Hall probes were arranged at the center and 4 mm away from the center of CCs as shown in Fig. 1(b). All these CCs and the Cu plate were compressed between the Cu lid and the Cu casing by screws. In addition, a Cernox thermometer was attached in the Cu lid to measure the variation of temperature of the stacked magnet. The schematic illustration of the completed stacked magnet assembly is shown in Fig. 1(a), and the photo is shown in Fig. 1(c).

For the evaluation of $J_c$ characteristics of CCs, magnetization measurements up to 5 T were conducted by magnetic property measurement system (MPMS, Quantum Design) on a small piece of CC that was carefully cut by using a wire saw, and $J_c$ was calculated by using the extended Bean model [24]. CCs used for the stacked magnet were cut by a mechanical cutter (shearing machine). In order to inspect possible damages in the cutting process, magneto-optical (MO) images were taken on a small piece cut by both the mechanical cutter and wire saw. It should be noted that the time needed to make a single cut is ~10 s using the mechanical cutter, while it is 15-30 min using the wire saw. MO images are obtained using an indicator film, which is a ferrite garnet film grown by the liquid epitaxy on a gadolinium gallium garnet substrate. The sample is cooled with a He-flow cryostat (MicrostatHighRes II, Oxford Instruments). A cooled-CCD camera (ORCA II-ER, Hamamatsu) and an objective lens (MPlan20x, Nikon) are used to acquire images. Magnetic images of the remanent state after applying positive and negative magnetic fields of $\mu_0 H = 0.1$ T are captured and subtracted to construct differential images. More than ten differential images are averaged to obtained the final MO image with improved magnetic resolution. For trapping magnetic field, an 18 T superconducting magnet at Institute for Materials Research (IMR, Tohoku University) was used, and magnetization was conducted by field-cooling method. The trapped magnetic field was measured by passing 0.1 mA current through the Hall probe by current sources (7651, YOKOGAWA), and voltages were measured by digital volt meters (34420A, Agilent). To eliminated uncertainty in the Hall resistance from the offset, each point was measured for positive and negative currents. To convert the Hall resistance to the value of trapped magnetic induction, we made separate calibrations of the Hall probe without the sample at several temperatures. Thus obtained calibration curves are included in the Supplemental Information.

## 3. Results and Discussion

*3.1 Characterizations of coated conductors*

Figures 2(a) and (b) show MO images of CCs in the remanent state at 85 K cut with a wire saw and with a mechanical cutter, which was adopted as the cutting method to fabricate the stacked magnet, respectively. The original long directions of the CC are shown by double-headed arrows. It is interesting to note that complete roof-top profiles are distorted by the presence of lower magnetic field (dark contrast) regions running predominantly along the vertical direction in both cases. Regions with low magnetic inductions are flux jets, where the magnetic induction profile is strongly distorted by the presence of defects in superconductors and the distortion is amplified by the strong nonlinearity of the $E-J$ characteristics of superconductors [25,26]. Almost no defects are observed near the vertical edge of the CC cut by the wire saw. The strongest distortion of flux profile is observed near the horizontal edge in the CC cut by the mechanical cutter. The presence of these defects makes the average $J_c$ along the horizontal direction lower. It can be quantified by the angle of the current-discontinuity line (*d*-line) with respective to the edge of the sample for a nearly rectangular sample. The two angles shown in Fig. 2(b) are $\theta_1 = 51°$ and $\theta_2 = 39°$, so the estimated anisotropy in $J_c$ is about 1.23. Since the anisotropy in $J_c$ induced by defects are mostly confined to regions close to the edge cut with the mechanical cutter, the distortion of the *d*-line becomes smaller inside the sample, and the slope approached to 45°. On the other hand, the two angles shown in Fig. 2(a) are $\theta_1 = 49°$ and $\theta_2 = 41°$, so the estimated anisotropy in $J_c$ of the sample cut by the wire saw is about 1.15. Obviously, significant amount of defects are introduced by cutting the CC using the mechanical cutter. Fig. 3 compares $J_c - H$ characteristics of CCs cut by the two different methods using the wire saw and mechanical cutter. $J_c$ of the CC cut by the mechanical cutter is about 40% smaller than that cut by the wire saw. This large difference in $J_c$ of the two CCs cut by different methods can be understood as follows. As is evident from Fig. 2(b), defect region in CC cut by the mechanical cutter extends ~0.5 mm from the horizonal edge. Since the dimensions of the CC used for the $J_c$ measurements is 1.2×1.3 mm$^2$, it is expected that a large portion of the CC is affected by the defects introduced by the mechanical cutter. In addition, defects introduced from vertical edges also help to reduce the average $J_c$ slightly in this CC. In this respect, cutting CCs using the mechanical cutter is not ideal. However, the time needed to cut CCs using the wire saw is very much longer than that using the mechanical cutter. So, we employed the mechanical cutter for cutting CCs to be used for the stacked magnet. It should be noted that defects other than those described above were not observed, and significant distortion of the *d*-line was not observed, when the mechanical cutter is used for cutting.

*3.2 Trapped field*

For trapping magnetic field in the stacked magnet, it was field-cooled under the external field of 18 T from 100 K to the target temperature. After the temperature was stabilized, the external field was ramped down at a specific ramp rate. Fig. 4 shows the magnetic field, *B*, at two locations measured by the Hall probes and temperature measured by the thermometer illustrated in Fig. 1(a), when the external magnetic field was ramped down to zero at a rate of (a) 0.4 T/min and (b) 0.2 T/min. In both cases, when the external field was ramped down, flux jumps occurred, and the trapped field was decreased to smaller values very quickly. At the same time, temperature of the stacked magnet increased up to 35 K for a moment. Flux jumps occurred at an external field of $\mu_0 H = 1.8$ T at a ramp rate of 0.4 T/min, and $\mu_0 H = 0.5$ T at a rate of 0.2 T/min. Hence, we expected that flux jumps in the stacked magnet at 10 K can be suppressed by reducing the ramp rate a little bit more. Fig. 5 shows the magnetic fields and the temperature of the stacked magnet in the measurement at a ramp rate of 0.1 T/min. During the magnetization process, the temperature of the sample increased slightly but steadily, and the final temperature was 10.4 K. Finally, a magnetic field of 14.86 T was successfully trapped at the center of the stacked magnet. Trapping of magnetic field at other temperatures was also attempted, and the results are shown in Fig. 6 for the measurements at (a) 20 K and (b) 7 K. A field of 12.4 T was trapped at 20 K at a rate of 0.5 T/min. The external field was ramped down from 15 T in this measurement. So we succeeded in trapping a high magnetic field within 30 min. During the measurement at 7 K, we changed the ramp rate from 0.4→0.5→0.2→0.1 T/min, while the external field was ramped down from 18→15→8→3→0 T to save the magnetization time. The ramp rate of 0.4 T/min is the limit of the superconducting magnet when the field is changed from 18 T to 15 T. We failed to trap a high magnetic field at 7 K because a flux jump occurred even at a ramp rate of 0.1 T/min, at which rate we succeeded in trapping a high field at 10 K. It should be noted that even though a flux jump occurred at $\mu_0 H = 1.3$ T at 7 K, there were no flux jumps at high external fields even at 0.1 T/min. So, it might be possible to trap larger magnetic fields at 7 K by reducing the ramp rate further below $\mu_0 H \sim 2$ T. We also conducted the magnetization of the stacked magnet at 30, 50, and 75 K. Solid circles and triangles in Fig. 7(a) show temperature dependences of measured trapped field at two locations. The temperature, the trapped field, and the ramp rate at each temperature are summarized in Table 1. The trapped field increases when the temperature is decreased since $J_c$ increases as shown in Fig. 7(b). However, the behavior of temperature dependence of the trapped field is different from that of $J_c$. It could partly be due to the fact that $J_c$ decreases strongly at high fields. Here, $J_c$ values at $\mu_0 H =$ 10 T, 15 T, and 20 T are extrapolated values from measured $J_c - H$ curve by using the following equation [27];

$$J_c(H) = C \frac{F^{\max}}{H^{\max}} \left(\frac{H + H_0}{H^{\max}}\right)^{p-1} \left(a - \frac{H + H_0}{H^{\max}}\right)^q, \tag{1}$$

where $F^{\max}$ is the maximum flux pinning force density, $H^{\max}$ is the magnetic field when pinning force density is maximum, $H_0$ is a magnetic field constant to avoid a divergence of $J_c$ at low magnetic fields, and $a$, $C$, $p$, and $q$ are parameters of the scaling function.

Table 2 summarizes characteristics of our trapped-field magnet and other magnets with high trapped fields at similar temperatures. Trapped fields of bulk superconductors are evidently higher than stacked magnets because of larger engineering current density. On the other hand, ramp rates to achieve high trapped field in bulk superconductors are low compared with those for stacked magnets. Compared with Ref. [18], the trapped field of our stacked magnet is about 95%, while the volume and the number of stacked CCs are much smaller. Compared to Ref. [17], while the volume is the almost same, the trapped field of our stacked magnet is 10% higher, and the number of stacked CCs are only a half. This indicates that the CCs used in the present study have great characteristics for materials to be used in stacked magnets. Furthermore the ramp rate of our stacked magnet is much higher than other magnets because Ag layer on the CCs and strong compression of CC stacks yield good thermal stability.

In the real application of the stacked magnet, the stability of the trapped field is also important. So, the relaxation of the trapped field was measured for 20 minutes after the magnetization of the stacked magnet. Fig. 8 shows the result of the measurement of flux creep at 10 K. The normalized relaxation rate ($d \ln M / d \ln t$) is estimated as 0.012. This value is close to the relaxation rate of REBCO film with artificial pinning centers of 0.016-0.038 [28,29]. It can be estimated that more than 90% of trapped field would be maintained even 24 hours after the magnetization.

**Table 1.** Temperature, trapped field, and ramp rate at each temperature.

| $T$ (K) | 10.4 | 19.8 | 31.2 | 50.6 | 74.9 |
|---|---|---|---|---|---|
| $B_{\text{trap}}$ (T) | 14.8 | 12.4 | 9.76 | 5.94 | 1.65 |
| Ramp rate (T/min) | 0.1 | 0.5 | 1 | 1 | 1 |

Table 2. Characteristics of several trapped-field magnets.

| Magnet | Present study | Ref. [18] | Ref. [17] | Ref. [2] | Ref. [6] |
|---|---|---|---|---|---|
| Type | stack | hybrid stack | stack | bulk | bulk |
| Number of CCs | 200 | 822 | 400 | - | - |
| $T$ (K) | 31 | 30 | 28 | 26 | 30 |
| $B_{\text{trap}}$ (T) | 9.8 | 10.4 | 8.6 | 17.6 | 16.9 |
| Volume (cm$^3$) | 1.9 | 33.8 | 2.0 | 13.7 | 21.7 |
| Ramp rate (T/min) | 1 | 0.05 | 0.15-0.3 | 0.015 | 0.05 |

*3.3 Numerical Calculation*

In order to obtain a quantitative understanding of the measured trapped field, the magnetic field profile of the trapped field in the central plane of the stacked magnet was calculated. The dimensions of the stacked magnet was assumed to be 12×12×12.5 mm$^3$ in the calculation based on the presence of defects as described in section 3.1. The reduction of sample dimensions plays the same role as the anisotropy of $J_c$ for the calculation of the trapped field because it is determined by $I_c = J_c \times$ (sample dimension). In our calculation, since the reduction of the dimensions is easier to handle, we assumed isotropic $J_c$ and reduction of sample dimensions in this section. The procedure is the same as the method reported previously [19]. Namely, with fully taking into account the local $J_c - H$ characteristics, the magnetic field profile was calculated by numerical integration based on Biot-Savart law. To increase the accuracy of the calculation, we divided each CC layer into 60×60 mesh as compared with 20×20 in the previous calculation. $J_c(B)$ data shown in Fig. 3 were fitted by eq. (1) for $\mu_0 H > 1$ T and eq. (2) shown below [30,31] for $\mu_0 H < 1$ T, and the fitting curve was used for the calculation;

$$J_c(H) = J_{c1} \exp\left(-\frac{H}{H_L}\right) + J_{c2} \frac{H}{H_{\max}} \exp\left[\frac{1}{\alpha}\left(1 - \left(\frac{H}{H_{\max}}\right)^\alpha\right)\right], \tag{2}$$

where $J_{c1}$ and $H_L$ are the height and field scale of the center peak, respectively, and $J_{c2}$ and $H_{\max}$ are the height and position of the second peak, respectively with $\alpha$ being a parameter characterizing the exponential decay of the second peak at high fields. Fig. 9(a) shows the comparison between the measured trapped fields and the calculated profile at temperatures between 10 K and 30 K. The experimental values agree very well with the calculated profile at 20 K and 30 K. However, the experimental values are obviously smaller than the calculated profile at 10 K. In Fig. 7(a), open circles and triangles show temperature dependences of the calculated magnetic field at the locations of the Hall probes (center and 4 mm away from the center). It is obvious that the measured magnetic field deviates from the calculated lines at 10 K. Since flux jumps prevented trapping a large magnetic field below 10 K, the smaller value of the trapped field at 10 K can be related to smaller-scale flux jumps and/or flux jets induced by the defects close to the edge of CCs. The length of flux jets, $L$, perpendicular to current is given by $L \sim na$, where $n$ is a $n$-value of $E-J$ characteristics and $a$ is an actual size of the defect. Generally, the $n$-value becomes larger at lower temperatures, leading to larger contribution of flux jets at lower temperatures. Hence, the dimensions of the stacked magnet is effectively reduced due to the presence of defects, leading to smaller trapped field at low temperatures. In order to evaluate the effective dimensions of CCs at 10 K, several calculations for different lengths of CCs were performed. Fig. 9(b) shows the plots of the error $(B_{\text{cal}} - B_{\text{exp}})^2 / B_{\text{exp}}^2$ vs the length of the CC along $x$-direction and the quadratic fitting. From this fitting, we can conclude that the effective length of CCs is reduced to 10.17 mm at 10 K. The calculated trapped field for stacked magnet with dimensions of 10.17×12×12.5 mm$^3$ is 14.79 T, which agrees very well with the experimental value of 14.8 T.

        Finally, we propose the prospect of increasing the trapped field of our stacked magnet. From the comparison between the experiment and the calculation, we found that the stacked magnet does not make use of full geometrical dimensions of the CCs due to the reduction of the effective dimensions. This reduction is caused by flux jets, which are originated from defects produced in the cutting process. So, we can suppress the reduction of the effective dimensions by employing cutting methods which induce fewer and/or smaller flux jets like in Fig. 2(a). In addition, flux jets can play a role to induce flux jumps. So, we can trap a high magnetic field at temperatures lower than 10 K using the sample with suppressed flux jets.

## 4. Summary

We have fabricated a stacked magnet by stacking 200 pieces of EuBCO coated conductors with BaHfO$_3$ nanorods, and it successfully trapped a field of 14.8 T at 10 K after ramping down the external field at a rate of 0.1 T/min. This value is the highest ever reported for the stacked CCs with a width $\leq$ 12 mm. The calculation of trapped field was also performed to understand the experimental results quantitatively. The calculated value is a little larger than the experimental value at 10 K even though they agrees well at higher temperatures. We interpret the reduced trapped field by the reduction of the effective dimensions of the stacked magnet due to the influence of defects observed in MO image, which becomes more significant at low temperatures.

**Acknowledgements**

A part of this study was performed at the High Field Laboratory for Superconducting Materials, Institute for Materials Research, Tohoku University (Project No. 19H0069).

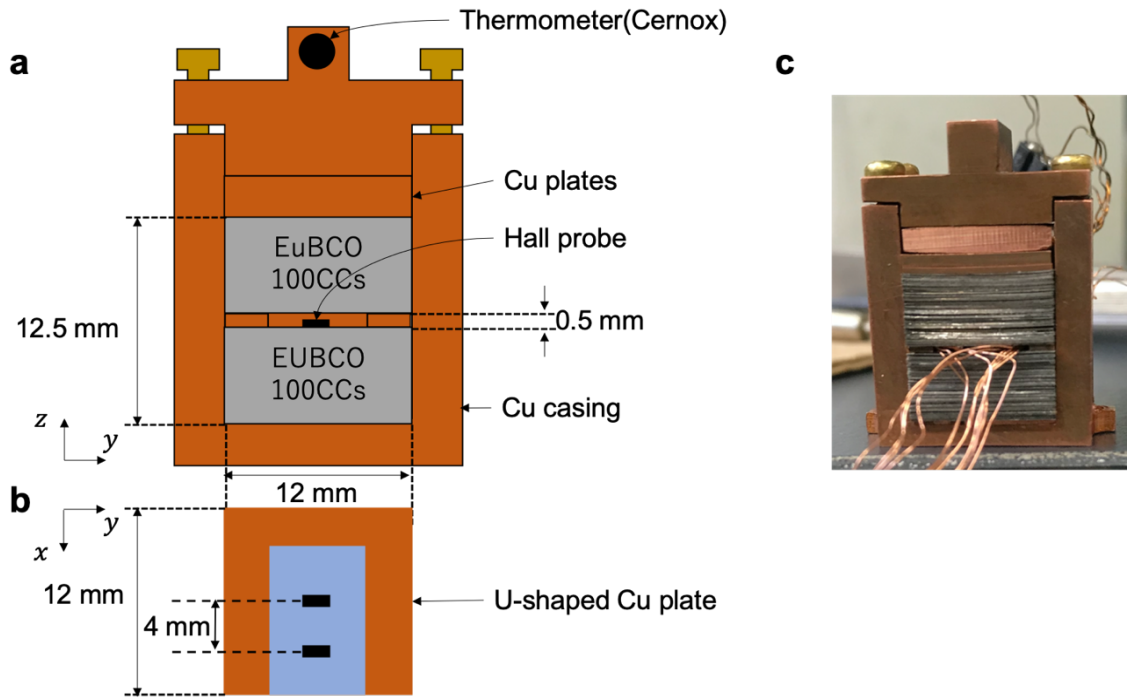

**Figure 1.** (a) Schematic illustration of stacked CCs and other components. (b) Arrangements of Hall probes in the Cu plate between the two 100 CCs. (c) A photo of stacked CCs in a Cu casing.

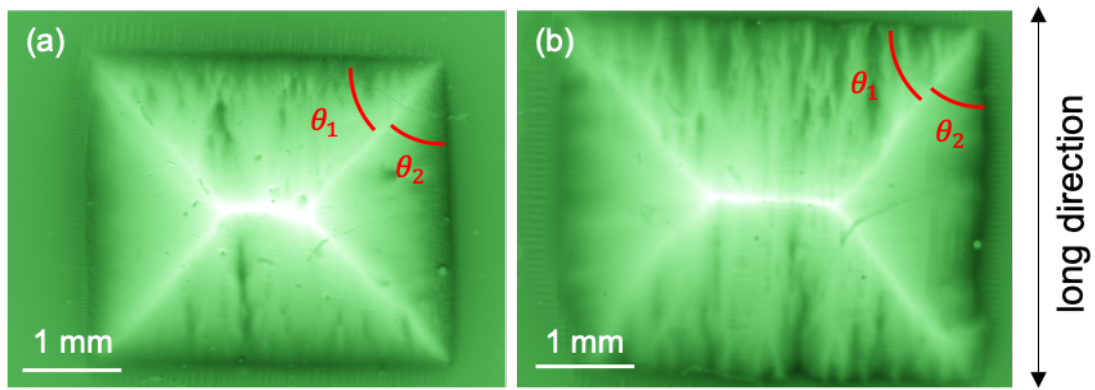

**Figure 2.** Magneto-optical images of small pieces of CCs in the remanent state at 85 K cut using (a) a wire saw and (b) a foot-operated mechanical cutter.

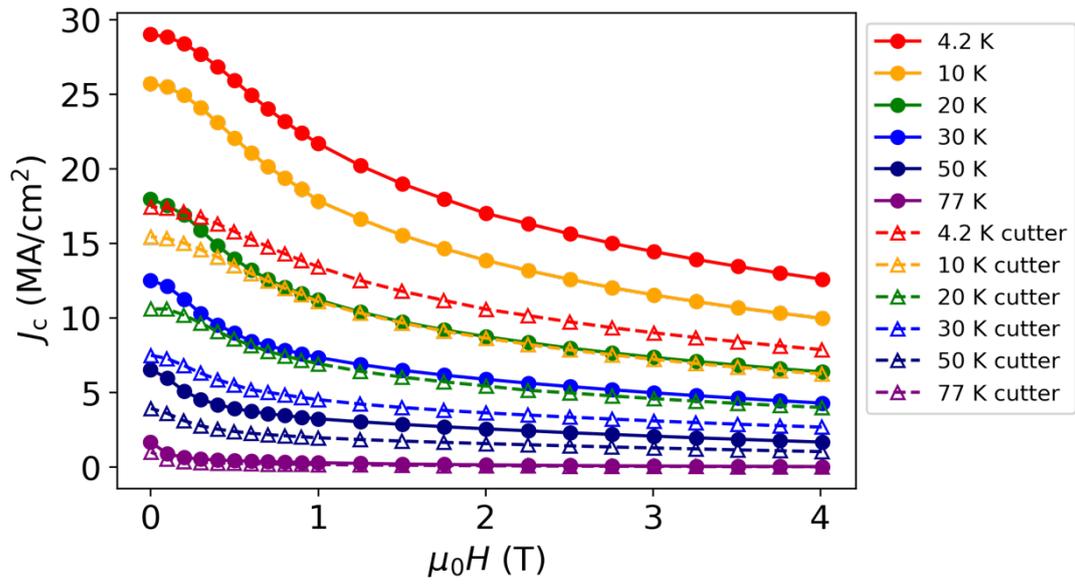

**Figure 3.** Magnetic field dependences of $J_c$ for small pieces of CCs at various temperatures. Solid and open symbols are data for CCs cut using a wire saw and a mechanical cutter, respectively.

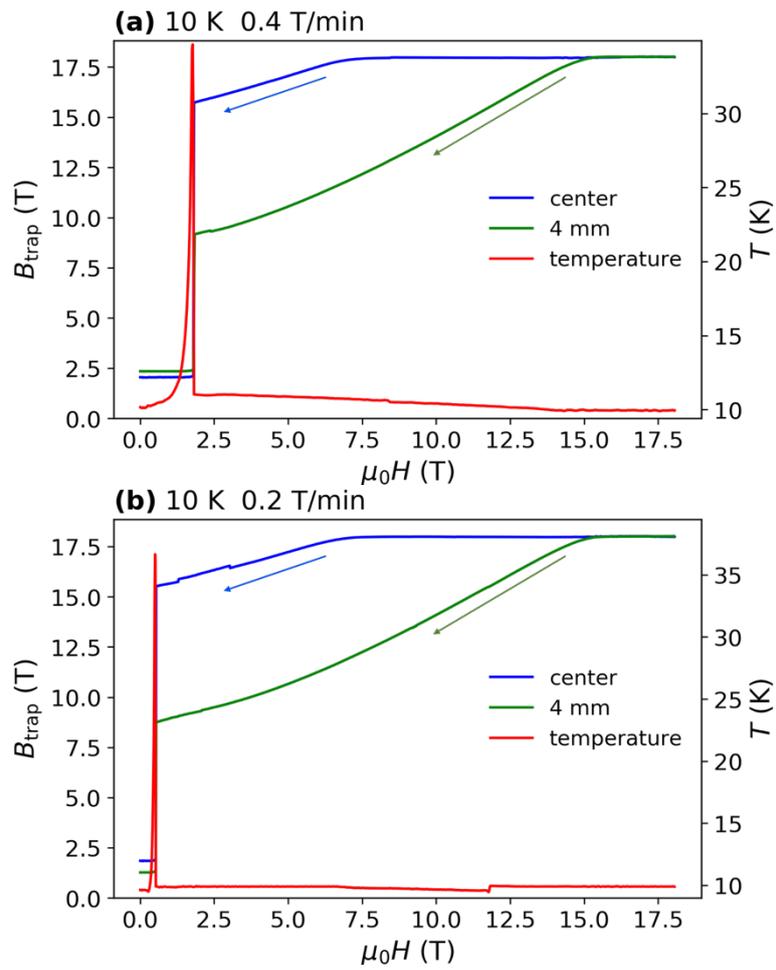

**Figure 4.** Magnetic fields at two locations and temperatures of the stacked CCs when the external magnetic field was ramped down at a rate of (a) 0.4 T/min and (b) 0.2 T/min.

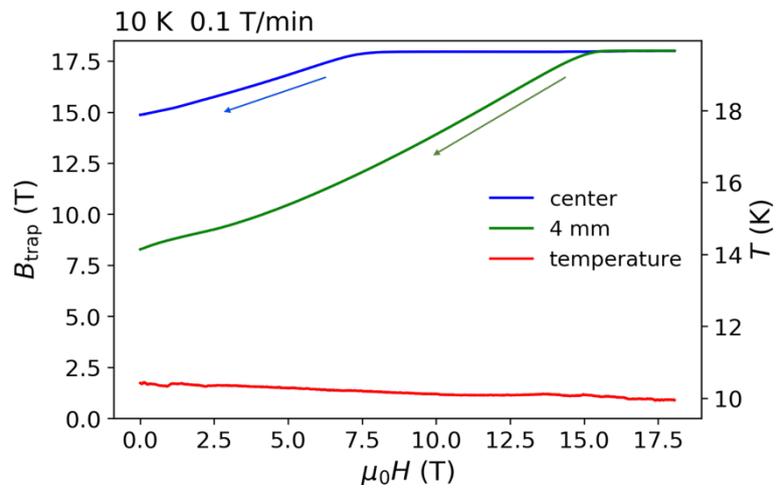

**Figure 5.** Magnetic fields measured at two locations of the stacked CCs when the external magnetic field was ramped down from 18 T at a rate of 0.1 T/min. Temperature measured near the stacked CCs is also shown.

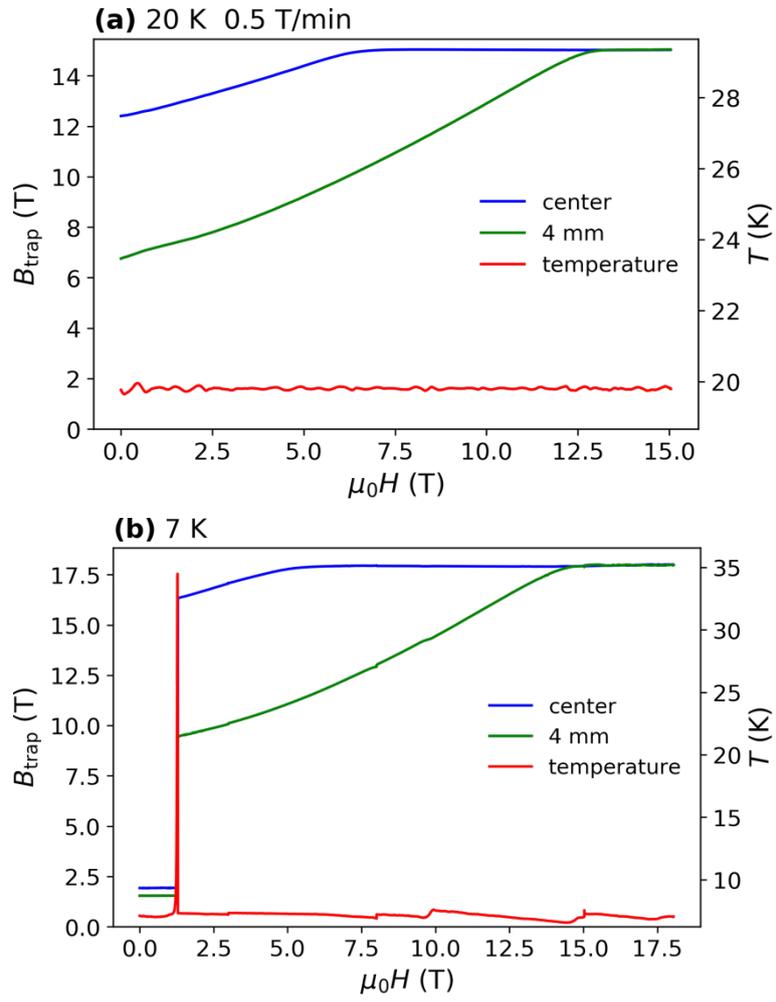

**Figure 6.** Magnetic fields at two locations and temperatures of the stacked CCs when the external magnetic field was ramped down at (a) 20 K and (b) 7 K.

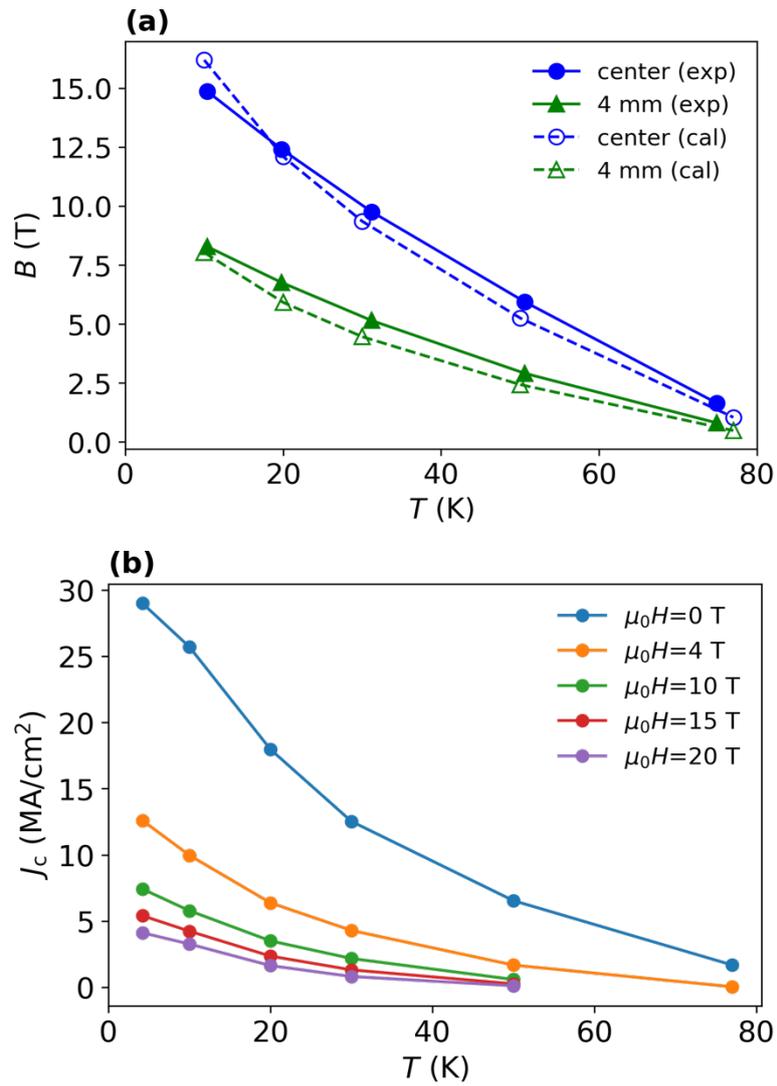

**Figure 7.** Temperature dependences of (a) trapped field at two locations and (b) $J_c$ at various fields. The plots for $\mu_0 H = 10$ T, 15 T, and 20 T are extrapolated values.

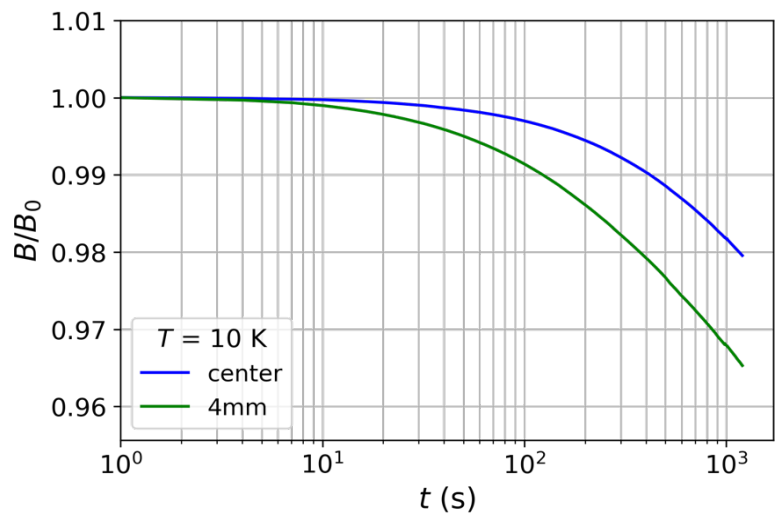

**Figure 8.** Temporal evolutions of normalized trapped fields measured at two locations of the stacked CCs after magnetization at 10 K.

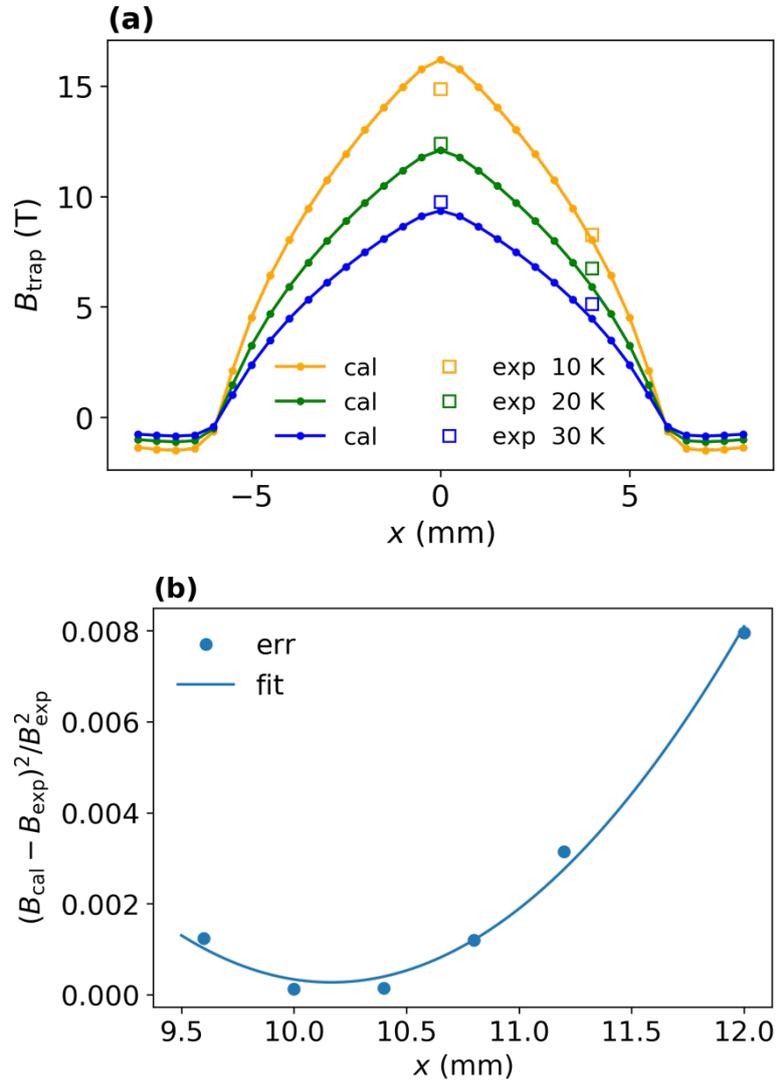

**Figure 9.** (a) Comparison between the measured trapped field at two locations and calculated magnetic field profile assuming the sample dimensions of 12×12×12.5 mm$^3$. (b) Errors between the measured trapped field at the center of the CC stacks and calculated field values assuming that the length of CCs is $x$ mm.

# Trapping a magnetic field of 14.8 T using stacked coated conductors of 12 mm width

Masahiro Suyama[1], Sunseng Pyon[1], Yasuhiro Iijima[2], Satoshi Awaji[3], and Tsuyoshi Tamegai[1]

[1] Department of Applied Physics, The University of Tokyo, 7-3-1 Hongo, Bunkyo-ku, Tokyo 113-8656, Japan
[2] Fujikura Ltd., 1-5-1 Kiba, Koto-ku, Tokyo 135-8512, Japan
[3] Institute for Materials Research, Tohoku University, 2-1-1 Katahira, Aoba-ku, Sendai 980-8577, Japan

1. **Calibration of the Hall probe**

In order to convert the Hall resistance to the value of trapped magnetic field, we made calibrations of the Hall probe at several temperatures. The set-up for the measurements is almost the same as described in Sec. 2, and the only difference is that we arranged aluminum plates instead of EuBCO CCs. Fig. S1 shows the magnetic field dependences of Hall resistance of the two Hall probes at the center and 4 mm away from the center at several temperatures. The values of the resistance of two Hall probes at the same magnetic field are different, but normalized curves overlap to each other. So, we can use this calibration curve for the same type of Hall probes (HG-0711, Asahi Kasei Microdevices) universally.

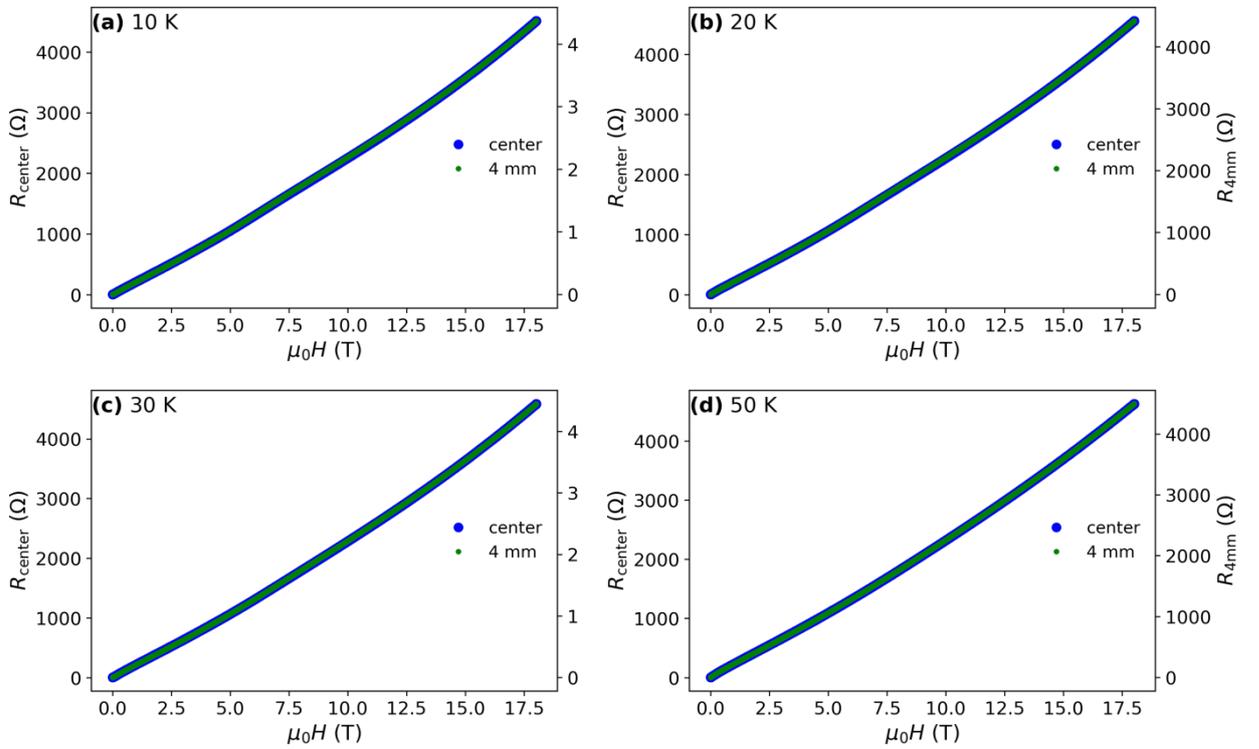

**Figure S1.** Magnetic field dependences of Hall resistance of the two Hall probes at the center ($R_{\text{center}}$) and 4 mm away from the center ($R_{\text{4mm}}$) at (a) 10 K, (b) 20 K, (c) 30 K, and (d) 50 K.

## 2. Relaxation of the trapped field at high temperatures

After the magnetization of the stacked magnet, temporal evolutions of the trapped field were measured for 20 min. In Sec. 3.2, the temporal evolution of the normalized trapped field at 10 K is shown. As additional data, temporal evolution of the normalized trapped field at 20 K. 30 K, 50 K, and 75 K are shown in Fig. S2.

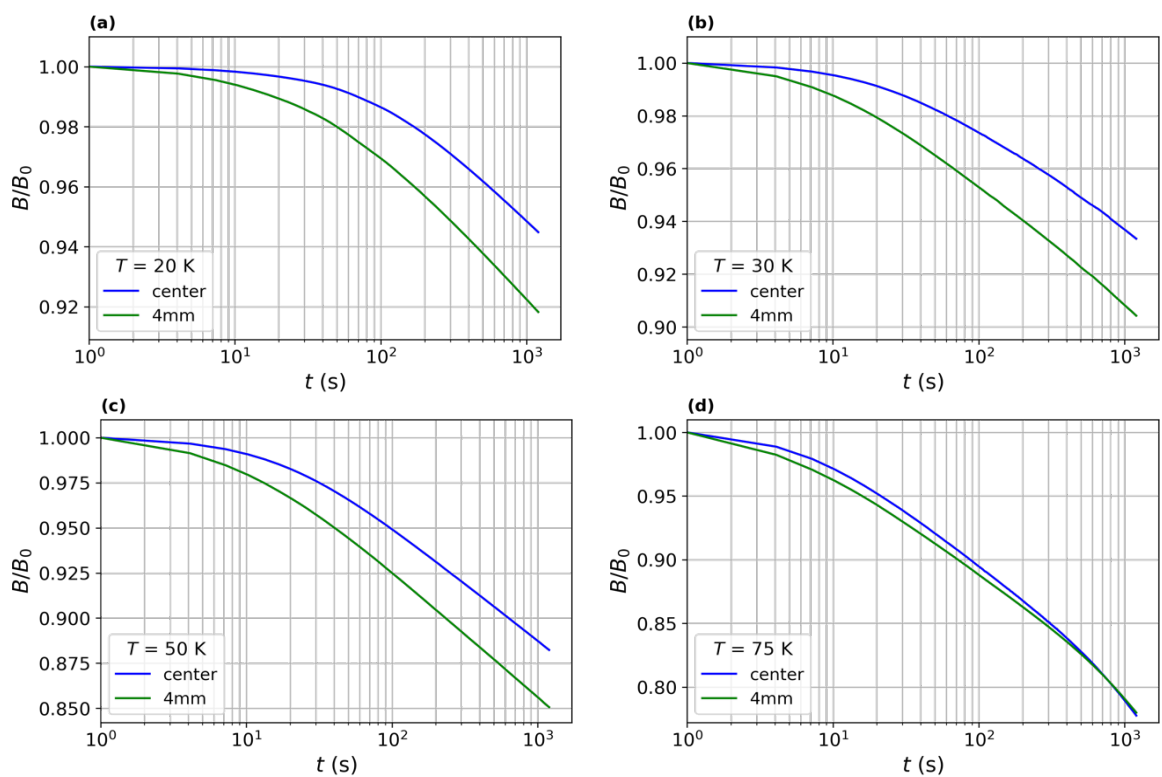

**Figure S2.** Temporal evolutions of the normalized trapped field measured at two locations of the stacked CCs after magnetization at (a) 20 K, (b) 30 K, (c) 50 K, and (d) 75 K.